\documentclass[aps,pre,twocolumn,amsmath,amssymb,superscriptaddress,floatfix]{revtex4}

\usepackage{graphicx}
\usepackage{bm}
\usepackage[usenames]{color}
\bibstyle{apsrev.bib}

\newcommand{\be}{\begin{equation}}
\newcommand{\ee}{\end{equation}}
\newcommand{\beqn}{\begin{eqnarray}}
\newcommand{\eeqn}{\end{eqnarray}}

\begin{document}

\title{Scaling of local persistence in the disordered contact process}
\author{R\'obert Juh\'asz}
\email{juhasz.robert@wigner.hu}
\affiliation{Wigner Research Centre for Physics, Institute for Solid State Physics and Optics, H-1525 Budapest, P.O.Box 49, Hungary}

\author{Istv\'an A. Kov\'acs}
\email{istvan.kovacs@northwestern.edu}
\affiliation{Department of Physics and Astronomy, Northwestern University, 2145 Sheridan Road, Evanston, IL 60208-3112, USA}
\affiliation{Wigner Research Centre for Physics, Institute for Solid State Physics and Optics, H-1525 Budapest, P.O.Box 49, Hungary}

\date{\today}

\begin{abstract}
We study the time-dependence of the local persistence probability during a non-stationary time evolution in the disordered contact process in $d=1,2$, and $3$ dimensions. We present a method for calculating the persistence with the strong-disorder renormalization group (SDRG) technique, which we then apply at the critical point analytically for $d=1$ and numerically for $d=2,3$. According to the results, the average persistence decays at late times as an inverse power of the logarithm of time, with a universal, dimension-dependent generalized exponent. 
For $d=1$, the distribution of sample-dependent local persistences is shown to be characterized by a universal limit distribution of effective persistence exponents.   
By a phenomenological approach of rare-region effects in the active phase, we obtain a non-universal algebraic decay of the average persistence for $d=1$, and enhanced power laws for $d>1$. As an exception, for randomly diluted lattices,  the algebraic decay holds to be valid for $d>1$, which is explained by the contribution of dangling ends.   
Results on the time-dependence of average persistence are confirmed by Monte Carlo simulations. 
We also prove the equivalence of the persistence with a return probability, a valuable tool for the argumentations.

\end{abstract}

\maketitle

\section{Introduction}

The local persistence in nonequilibrium systems attracted a lot of attention, as it reveals deep insights about the nonequilibrium dynamics and it often shows a nontrivial behavior \cite{bms,majumdar,redner}.
Generally, persistence is defined as the probability that a local field does not cross a given level up to time $t$. 
A simple example is a random walk on a line, where persistence measures the probability that the walker does not pass the origin up to time $t$. 
In systems with many degrees of freedom, the persistence probability typically has a power-law temporal decay, 
\be 
P(t)\sim t^{-\Theta},
\label{pl} 
\ee
where the exponent $\Theta$ is often non-trivial, even in simple systems such as a diffusive field with a random initial condition \cite{msbc,dhz,schehr}.  
Exact results in interacting systems are scarce, an exception is the persistence exponent of the one-dimensional, zero temperature $q$-state Potts model \cite{derrida}. 

Similar power-law behavior can be observed in critical nonequilibrium models belonging to the directed percolation (DP) universality class \cite{odor,hhl}, such as the contact process \cite{cp,liggett,md}. 
Here, the persistence $P(t)$ can be defined as the probability that, starting the system in a finite-density (non-stationary) state, an initially inactive site is not activated until time $t$. 
The persistence exponent can be regarded as a critical exponent which is independent of the standard critical exponents. According to simulations, $\Theta$ is universal and depends only on the dimension, up to the upper critical dimension $d_u=4$ for several models in the DP class \cite{hk,am,menon}; for exceptions see Refs. \cite{matte,saif}. Above the upper critical dimension, $\Theta$ is nonuniversal (model-dependent) \cite{fuchs,grassberger}. 
In the interpretation of the contact process as an epidemic spreading model, $P(t)$ is nothing but the probability that an individual is not infected until time $t$, a natural quantity to study.

The behavior of persistence in systems with quenched disorder is comparatively less known. One example for which exact results exist, is a one-particle problem, the random walk in random environment \cite{rwre}, which, in the case of continuous space and time, is also known as Sinai model \cite{sinai}. 
According to exact results \cite{comtet,mc}, the average persistence in the recurrent (driftless) Sinai model follows the law  
\be 
\overline{P(t)}\sim (\ln t)^{-\overline{\Theta}}
\label{log}
\ee
at late times, with $\overline{\Theta}=1$. 
The same logarithmic scaling was found by a strong-disorder renormalization group (SDRG) method \cite{fdm1,fdm2}, and also for the lattice variant of the model \cite{ir}.
Thus, in this example, quenched disorder changes the power law, frequently appearing in homogeneous systems, to a logarithmic scaling. This type of logarithmic dynamical scaling is typical in systems, where the critical behavior is controlled by an infinite-disorder fixed point (IDFP) of the SDRG transformation \cite{im}. Besides the Sinai model, another example of this class of models is the disordered contact process (DCP) \cite{hiv,moreira}. 
In this work, we aim at studying the persistence probability in the DCP on one, two, and three-dimensional lattices by means of phenomenological scaling, the SDRG method, and Monte Carlo (MC) simulations. 
Recently, a similar, one-dimensional model which is a quenched mixture of sites obeying rules of DP class and compact DP class has been considered in Ref. \cite{bg}, and an active Griffiths phase with varying, nonuniversal persistence exponents has been observed in MC simulations. 
We point out similar Griffiths effects in the active phase of the DCP, although, in dimensions $d>1$, the average persistence is found to decay according to an enhanced power law in case of random rates, whereas, on diluted lattices, it keeps obeying power laws. In addition to this, at the critical point, the average persistence is found to follow a logarithmic decay given in Eq. (\ref{log}). According to our results, the generalized persistence exponent $\overline{\Theta}$, which is determined analytically for $d=1$ and estimated numerically for $d=2,3$, is universal, i.e. it is independent of the form of disorder. 

The rest of the paper is organized as follows. The model and the persistence probability are defined in Sec. \ref{sec:model}. Sec. \ref{sec:phen} is devoted to the phenomenological description of rare-region effects in the active phase. In Sec. \ref{sec:sdrg}, the SDRG approach to the calculation of the persistence is presented and applied analytically to the one-dimensional DCP and numerically in $d=2$ and $d=3$ dimensions. The results are confronted with Monte Carlo simulations in Sec. \ref{sec:MC}, and discussed in Sec. \ref{sec:discussion}. An exact  reformulation of the local persistence as a return probability, which will be used frequently, is derived in Appendix \ref{app:return}.

\section{The model}           
\label{sec:model}

The contact process is a continuous-time Markov process on a set of binary variables $n_i=0,1$ sitting at the sites of a $d$-dimensional hypercubic lattice \cite{cp,liggett,md}. Sites with $n_i=1$ ($n_i=0$) are called active (inactive). 
There are two kinds of transitions which take place independently. First, site $i$, provided it is active, becomes spontaneously inactive with a rate $\mu_i$. Second, an active site ($i$) activates its inactive nearest neighbors ($j$) with a rate $\lambda_{ij}$.  
In the disordered contact process, either the deactivation rates or the activation rates or both of them are independent, identically distributed random variables. The model shows a continuous phase transition at a critical value value of the control parameter $\Delta=\overline{\ln(\lambda/\mu)}$, above which the order parameter, the average density of active sites in the steady state, is non-zero, and zero otherwise. Here and in the following, the overbar stands for an average over the random rates.  

The local persistence probability in the homogeneous contact process is usually defined as follows \cite{hhl}. 
The system starts to evolve from an initial state with a density of active sites $\rho_0<1$, and the local persistence $P(t)$ is the fraction of lattice sites which are not once activated till time $t$. The initial state can either be an uncorrelated state or a state evolved from the fully active state up to some time $t_0$. The basic characteristics of $P(t)$ in the homogeneous model are the following. In the inactive phase it tends to a positive constant; in the active phase, $P(t)\to 0$ exponentially, while at the critical point, it vanishes according to a power law as given in Eq. (\ref{pl}).  

In the DCP, being not translationally invariant, it is reasonable to introduce the persistence probability $P_0(t)$ of a given site (which we will label by $0$) in a given realization of the random rates. 
For this, we assume that, at $t=0$, all but site $0$ is active. Then, $P_0(t)$ is the probability that site $0$ is not once activated till time $t$. 
Obviously, after some time has elapsed, the global density declines to some $\rho_0<1$, and one is up against a similar situation as assumed in the usually defined persistence $P(t)$. The average of $P_0(t)$ over disorder, $\overline{P_0(t)}$ has therefore the same late time behavior as $P(t)$, apart from the precise value of the prefactor which is of less importance. 

In the DCP, we are mainly interested in how the average persistence,  $\overline{P_0(t)}$ behaves at late times. As the average is independent of the choice of site $0$, we will ignore the label and simply write $\overline{P(t)}$.   Here, the main features are unaltered, i.e. it tends to a positive constant in the inactive phase, and zero otherwise, although, as it will turn out, the functional forms in the latter case are different compared to the homogeneous model. 

Special care is needed for a particular case of disorder, the random site dilution namely. In this case, a randomly selected fraction $c$ of lattice sites is deleted and thus unavailable for the activity. If $c$ is below the percolation threshold, the diluted lattice consists of a macroscopic component and a macroscopic number of finite-size fragments. Since the finite fragments can reach the absorbing (inactive) state in a finite time even in the active phase of the model, the initially inactive sites of such fragments can stay inactive forever with a nonzero probability. Therefore the average persistence would tend to a positive constant even in the active phase of the model. To avoid this trivial behavior in diluted lattices, we will ignore finite fragments, and consider the process on the macroscopic component only.

\section{Phenomenology in the active phase} 
\label{sec:phen}

It is well known that, in the inactive phase of the DCP, locally supercritical regions give rise to an anomalous, algebraic decay of the density \cite{noest}, which is analogous to Griffiths effects in quantum magnets \cite{gmc}. 

For similar reasons, the average persistence will have a slower-than-exponential decay in the active phase. 
Here, due to the disorder, even if the system is locally supercritical almost everywhere, it contains rare regions which are locally subcritical. 
Let us assume that these regions are compact, isotropic and are characterized by their radius $l$. The probability of occurrence of such regions of radius greater than $l$ is, with an exponential precision,  $P_>(l)\sim e^{-Al^d}$.
The persistence time of such a region, which is surrounded by an active background, is roughly given by the time $\tau$ the activity needs to penetrate to the center of the region. Since a rare region is locally subcritical, this time is exponentially large in the radius: $\tau\sim e^{Bl}$. The distribution of local persistence times has thus the large-$\tau$ tail
$P_>(\tau)\sim e^{-C[\ln(\tau/t_0)]^d}$, where $C=AB^{-d}$ and $t_0$ are nonuniversal positive constants. 
Under the assumptions made above, the average persistence 
can be calculated as $\overline{P(t)}\sim \int e^{-t/\tau}\rho(\tau)d\tau$, where $\rho(\tau)=-\frac{dP_>(\tau)}{d\tau}$ is the probability density of $\tau$. 
Using the saddle-point approximation to evaluate this integral for large $t$, we obtain 
\beqn 
\overline{P(t)}\sim e^{-C[\ln(t/t_0)]^d}   \quad (d>1) \nonumber \\
\overline{P(t)}\sim t^{-C-1}  \quad (d=1).
\label{epl}
\eeqn
Thus the average persistence is expected to decay in the active phase according to an enhanced power law for $d>1$, and to a power law with a nonuniversal exponent for $d=1$.    
Approaching the critical point, the characteristic linear size of rare regions diverges and the above theory breaks down. 

\section{Persistence by the SDRG method}
\label{sec:sdrg}

An efficient technique for studying the critical behavior of the DCP is the SDRG method \cite{hiv,im}, which is thus complementary to the phenomenological scaling presented in the previous section. 
It was first applied to the DCP with symmetric activation rates $\lambda_{ij}=\lambda_{ji}$ in Ref. \cite{hiv}. The SDRG method is a real-space renormalization procedure by which fast degrees of freedom are sequentially eliminated, resulting in a gradual decrease of the rate scale $\Omega=\max\{\lambda_{ij},\mu_i\}$, which is set by the maximal transition rate of the process. 
It consists of two kinds of local reduction steps. If the largest rate is an activation rate, $\Omega=\lambda_{ij}$, sites $i$ and $j$ form a cluster characterized by an effective deactivation rate: 
\be 
\ln \tilde \mu_{ij} = \ln \mu_i + \ln\mu_j - \ln\lambda_{ij} + \ln 2. 
\label{mu_rule}
\ee
When the largest rate is a deactivation rate, $\Omega=\mu_{i}$, site $i$ is eliminated and new interactions between all pairs ($j,k$) of its neighboring sites are generated with effective activation rates:    
\be 
\ln \tilde \lambda_{jk} = \ln \lambda_{ij} + \ln\lambda_{ik} - \ln\mu_i. 
\label{lambda_rule}
\ee
The critical behavior of the DCP is described by the infinite-disorder fixed point of the transformation, at which the distribution of logarithmic rates broadens without limits and the approximative reduction steps become asymptotically exact \cite{im}. 
As a result of this coarsening procedure, the lattice sites are arranged into a nontrivial set of practically noninteracting clusters, which are characterized by some effective deactivation rates, and the constituents of which are not necessarily adjacent on the lattice.      

To our knowledge, the SDRG method has not been applied for the calculation of persistence in the DCP so far. In the following, we describe how this can be captured by the SDRG technique.  
The tractability of persistence relies on the observation that the SDRG procedure mimics the time evolution of the DCP: Starting the process from a fully active state, the set of sites which are active with a high probability at some time $t$ are given within the SDRG by the set of clusters still active (i.e. not eliminated yet) at rate scale $\Omega=1/t$. 
Let us assume that site $0$ was initially inactive while all other sites were active. Obviously, if site $0$ is merged with another cluster in the course of the SDRG procedure, it looses its intactness with a high probability. This occurs when any of the activation rates connected to site $0$ is picked for decimation. 
There is, however, a difficulty here. Before this event could happen, site $0$ may be decimated out, and in this case, the procedure does not keep a record of the activation rates connected to site $0$ any longer. 
This problem can be avoided by the following modification. The deactivation rate of site $0$ is set initially to zero, $\mu_0=0$, which ensures that site $0$ is never decimated out. Note that this can be safely done since the persistence probability of site $0$ does not depend on $\mu_0$. 
With this modification, site $0$ will loose its persistence in the course of the SDRG precisely when it is merged with another cluster. 

We can arrive to the same conclusion by the following argument, as well. Let us consider a fixed realization of the DCP and a modified one, which differs from the original one in that the deactivation rate at site $0$ is set to zero, $\mu_0=0$. In the latter case, let $P_{\rm ret}^{(0)}(t)$ denote the probability that, starting the process with all but site $0$ inactive, the state at time $t$ returns to the initial state. 
As it is proved in Appendix \ref{app:return} by exploiting the duality property of the contact process \cite{schutz,hv}, this return probability precisely equals to the persistence probability $P_0(t)$ of site $0$:
\be
P_0(t)=P_{\rm ret}^{(0)}(t). 
\ee
We mention that a similar relationship is valid for another representant of the DP class, the bond directed percolation \cite{hk}, but, to our knowledge, this has not been proven for the contact process so far.  
In the SDRG picture, the condition of finding the modified process (where $\mu_0=0$) in the initial state (i.e. only site $0$ active) at time $t$, is that no other clusters are merged to site $0$ down to scale $\Omega=1/t$. This is the same condition we obtained above.  

\subsection{Average persistence in one dimension}

In one dimension, the calculation of the average persistence can be carried out analytically by the SDRG method. In fact, this is equivalent to the calculation of the surface magnetization of the random transverse-field Ising chain (RTIC), which is solved in Ref. \cite{fisher}.

To show this equivalence, we neglect the term $\ln 2$ in Eq. (\ref{mu_rule}), which can be safely done at the critical point, where the logarithmic rates increase without limits in the course of the SDRG \cite{sm}. This way, the renormalization scheme becomes formally identical to that of the RTIC.
First, let us consider a semi-infinite chain and study the persistence $P_{\rm surf}(t)$ of the first (surface) site. 
At the beginning of the SDRG procedure, the deactivation rate of this site is set to zero (and there is nothing to do with this site during the procedure anymore), and we are interested in the probability $Q_{\rm surf}(\Omega)$ that the first activation rate $\lambda_{01}$ is not decimated until the scale $\Omega$. 
This probability, when $\Omega=1/t$ is substituted in it, provides the time-dependence of the average persistence probability (of the surface site): 
$\overline{P_{\rm surf}}(t)\sim Q_{\rm surf}(\Omega=1/t)$. 

Due to the $\lambda\leftrightarrow\mu$ duality of the SDRG scheme in one dimension, which is salient comparing Eqs. (\ref{mu_rule}) and (\ref{lambda_rule}) (if the constant term is dropped), the probability $Q_{\rm surf}(\Omega)$ is the same as the probability of not decimating the surface site (or the cluster containing the surface site) of a semi-infinite system in which the distribution of activation rates and deactivation rates are interchanged. The evolution of this probability under the SDRG procedure has been calculated analytically in Ref. \cite{fisher} in the context of the RTIC, where it describes the scaling of the average surface magnetization. 
It turned out that critical systems with any initial distribution of rates flow towards a self-dual, universal IDFP, at which $Q_{\rm surf}(\Omega)\sim 1/\ln[\Omega_0/\Omega]$ \cite{fisher}. This yields for the time-dependence of the average persistence of the surface site
\be 
\overline{P_{\rm surf}}(t)\sim [\ln(t/t_0)]^{-1}.
\ee

In the case of the persistence of a bulk site of an infinite chain, there is no further complication. Site $0$ is connected to two semi-infinite chains, and we are interested in the probability $Q_{\rm bulk}(\Omega)$ that neither of the two activation rates connected to it are decimated down to scale $\Omega$. Obviously, until such an event the two halves of the system do not communicate, so we have simply $Q_{\rm bulk}(\Omega)=[Q_{\rm surf}(\Omega)]^2$. 
Therefore, we obtain for the average persistence of bulk sites at the critical point at late times: 
\be 
\overline{P}(t)\sim [\ln(t/t_0)]^{-2}.
\label{1d_bulk}
\ee

\subsection{Distribution of persistence in one dimension}

In the DCP, the persistence probability of a given site depends on the realization of disorder, thus it varies from sample to sample. Beyond the average over disorder, a complete characterization of persistence is given by the distribution $S(P_0,t)$, which depends on the parameter $t$.   
In one dimension, the SDRG scheme is simple enough so that the limit distribution at late times can be calculated as follows. 

To calculate the average it was a good (asymptotically correct) approximation to take the persistence after site $0$ was merged with another cluster as zero. 
Yet, to obtain the distribution, we must go beyond this point and take into account that, even after such an event, there remains a small but non-zero persistence probability. 
Let us consider the persistence of the first site of a semi-infinite chain, and describe it by a variable $p(\Omega)$ in the course of the SDRG. Initially, for $\Omega=\Omega_0$, it is set to one, $p(\Omega_0)=1$, and it will remain unchanged until the activation rate connected to it is decimated at some rate scale $\Omega_1$. This means that site $0$ is merged with the next cluster having a deactivation rate $\mu_1$. Since typically $\mu_1\ll\Omega_1$, the next cluster is active at time $t_1=1/\Omega_1$ with a probability close to one. However, it may have got deactivated by the time $t_1$, with a small probability $\mu_1t_1=\frac{\mu_1}{\Omega_1}$. In this case, the newly formed cluster containing site $0$ will be inactive, and site $0$ remains intact even after such an event. The variable $p(\Omega)$ is thus renormalized as 
\be 
\tilde p=p\frac{\mu_1}{\Omega_1}.
\label{p_rg}
\ee 
Similarly, a further merging of the surface cluster with the next one (having some deactivation rate $\mu_2$) at a lower renormalization scale $\Omega_2$, will reduce the variable $p$ by a factor $\frac{\mu_2}{\Omega_2}$. 

We can come to the same conclusion by using the equivalence of persistence with a return probability. In this case, the variable $p(\Omega)$ is interpreted as the return probability at time $t=1/\Omega$. It remains $1$ until the adjacent cluster with deactivation rate $\mu_1$ is merged with the surface site at scale $\Omega_1$. After this event, we have a new cluster with a simple internal dynamics. The surface site is always active (since $\mu_0=0$) while the other component can be deactivated with a rate $\mu_1$ and activated with a rate $\Omega_1$. The probability that only the surface site is active, which is nothing but the return probability to the initial state, is $\frac{\mu_1}{\Omega_1+\mu_1}\approx \frac{\mu_1}{\Omega_1}$. Thus, the variable $p(\Omega)$ transforms in the same way as given in Eq. (\ref{p_rg}).     

At the critical point, the transformation of the persistence probability in Eq. (\ref{p_rg}) is formally identical with that of the surface order parameter of the DCP, analyzed in Ref. \cite{juhasz_dist}, so we can make use of the results obtained there. As it is shown there, the fixed-point distribution of $K=\ln(1/p)$  is simply
\be 
B_{\Gamma}(K)=\frac{1}{\Gamma}e^{-K/\Gamma},
\ee
where $\Gamma=\ln(\Omega_0/\Omega)$. 
This yields that the persistence probability of a surface site has the distribution at late times
\be 
S_{\rm surf}(P_{\rm surf},t)=[\ln(t/t_0)]^{-1}e^{-\ln(1/P_{\rm surf})/\ln(t/t_0)}.
\ee  
The scaling variable $\Theta_{\rm s}=-\ln(P_{\rm surf})/\ln(t/t_0)$ appearing here can be interpreted as a sample and time-dependent effective persistence exponent, which has the limit distribution $\tilde S_{\rm surf}(\Theta_{\rm s})=e^{-\Theta_{\rm s}}$.  

In the case of a bulk site, the persistence probability $P_{\rm bulk}(t)$ is a product of two independent surface persistences corresponding to the two sides of  site $0$. The scaling variable $\Theta_{\rm b}=-\ln(P_{\rm bulk})/\ln(t/t_0)$ is thus a sum of two independent, exponentially distributed variables $\Theta_{\rm s}$, having the distribution
\be 
\tilde S_{\rm bulk}(\Theta_{\rm b})=\Theta_{\rm b}e^{-\Theta_{\rm b}}.
\ee
The result we just obtained can be interpreted that, as opposed to the homogeneous contact process, which is characterized by a single persistence exponent, the persistence in the critical DCP is described by an entire distribution of persistence exponents.  

Having the distribution of effective exponents, we can readily calculate other characteristics, such as the typical persistence defined as $[P_0(t)]_{\rm typ}=\exp\{\overline{\ln P_0(t)}\}$. For this, we obtain power-law decays, 
$[P_{\rm surf}(t)]_{\rm typ}\sim t^{-1}$ for a surface site and $[P_{\rm bulk}(t)]_{\rm typ}\sim t^{-2}$ for bulk sites.

\subsection{Higher dimensions}

In dimensions $d>1$, the SDRG method cannot be treated analytically. 
Here, we applied the numerical SDRG algorithm developed in Ref. \cite{kovacs}, which is very efficient in producing the final cluster structure of a finite sample at the expense of being agnostic of the decimation history. Yet, the method can be used to determine the time-dependence indirectly, through finite-size scaling, as follows.
In an ensemble of finite samples of linear size $L$, we calculated the fraction $P(L)$ of samples in which site $0$ (for which $\mu_0=0$) remained a one-site cluster in the final set of clusters.  
On the grounds of the critical scaling of $\overline{P}(t)$ in one dimension, we expect  
\be
\overline{P}(t)\sim [\ln(t/t_0)]^{-\overline{\Theta}}
\label{P_gen}
\ee
to hold with a dimension-dependent universal exponent $\overline{\Theta}(d)$, at least in dimensions $d<4$, where the validity of the SDRG approach is supported by Monte Carlo simulations \cite{vh}. 
Using the logarithmic dynamical scaling of the form $\ln(\Omega_0/\Omega)\sim L^{\Psi}$ valid at an IDFP \cite{im}, we obtain that the probability $P(L)$ must scale with the system size as   
\be 
P(L)\sim L^{-\Psi\overline{\Theta}}.
\ee 
By determining the exponent $x_p=\Psi\overline{\Theta}$ numerically, and using the known estimates $\Psi(d=2)=0.48(2)$ and $\Psi(d=3)=0.46(2)$, obtained in two \cite{kovacs} and three \cite{kovacs3d} dimensions by the numerical SDRG method, we can calculate the generalized persistence exponents.          

\begin{figure}[ht]
\includegraphics[width=8cm]{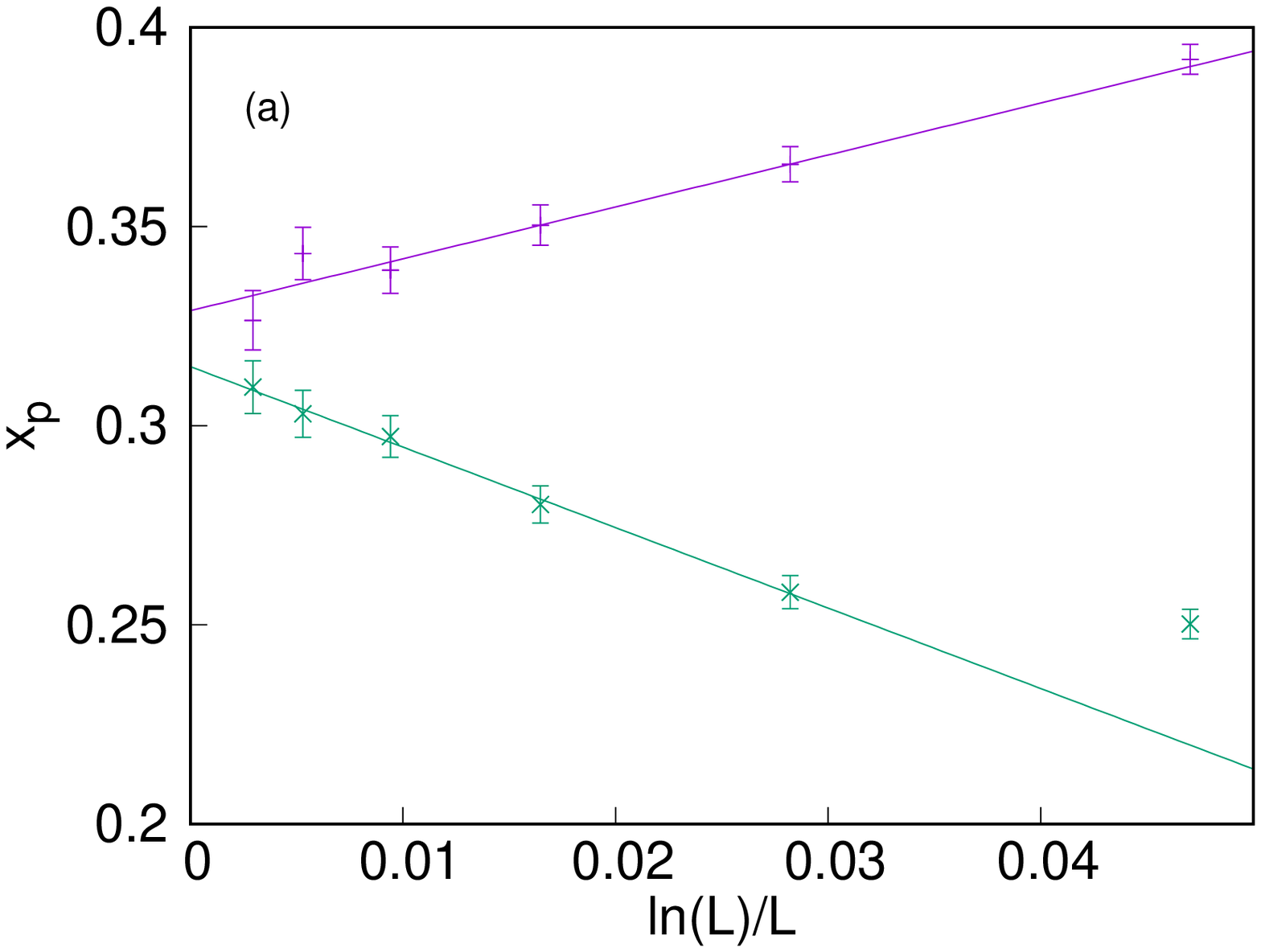}  
\includegraphics[width=8cm]{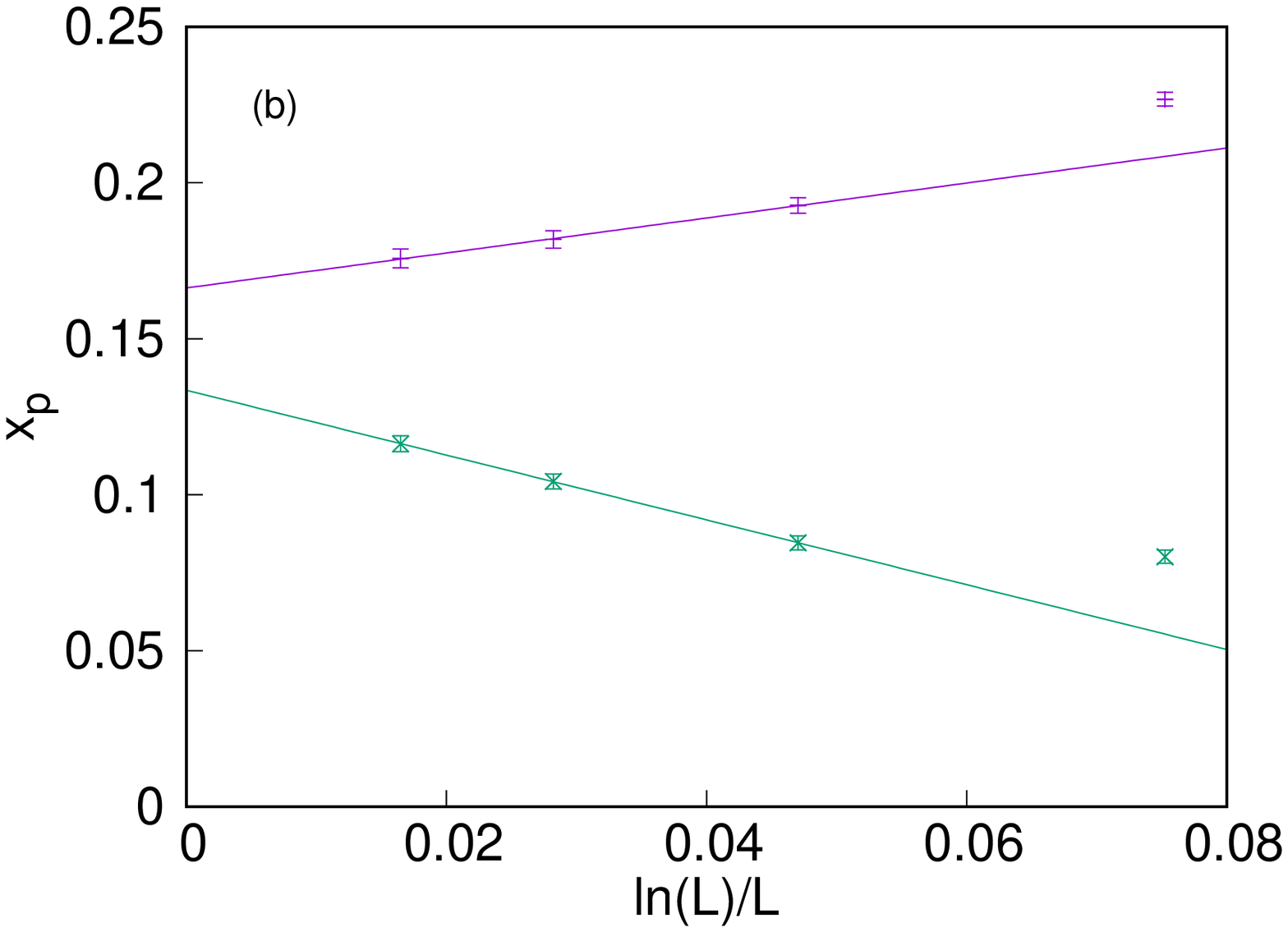}  
\caption{\label{rg} (Color online)  SDRG estimates of the exponent $x_p=\Psi\overline{\Theta}$ in the $d=2$ (a) and $d=3$ (b) critical model. The exponents are the results of two-point fits for size $L$ and $L/2$, providing consistent extrapolated values $x_p(d=2)=0.32(1)$ and $x_p(d=3)=0.15(2)$ as $L\to\infty$ for both box-$\mu$ (top, purple) and fixed-$\mu$ (bottom, green) disorder. The straight lines are linear fits to the data.  
}
\end{figure}

In our numerical calculations, we have renormalized finite samples of linear size up to $L = 1024$ and $L=128$ for $d=2$ and $d=3$, respectively, with periodic boundary conditions. The number of samples was at least $10^6$. To highlight the universality of the results, we have used two different parameter distributions to implement the disorder. The activation rates were chosen uniformly from the interval $\lambda_{ij} \in (0,1]$ in both cases, while the deactivation rates were either chosen from a uniform interval as $\mu_i \in (0,\mu]$ (box-$\mu$ disorder), or kept constant $\mu_i=\mu, \forall i$ (fixed-$\mu$ disorder) with a control parameter $\Delta = \ln(\mu)$. The location of the critical point is known for both disorder distributions to be at $\Delta_c^b(d=2)=1.6784(1)$, $\Delta_c^f(d=2)=-0.17034(2)$ \cite{kovacs} and $\Delta_c^b(d=3)=2.5305(10)$, $\Delta_c^f(d=3)=-0.07627(2)$ \cite{kovacs3d}. We find consistent, universal exponents for both disorder distributions, providing the $L\to\infty$ extrapolated exponents $x_p(d=2)=0.32(1)$ and $x_p(d=3)=0.15(2)$, see Fig. \ref{rg}, which yield the following estimates for the generalized persistence exponents:
\be 
\overline{\Theta}(d=2)=0.67(5), \quad \overline{\Theta}(d=3)=0.33(6).
\label{theta_sdrg}
\ee

\section{Monte Carlo simulations}
\label{sec:MC}

In order to check the results obtained by the phenomenological considerations and the SDRG method, we performed numerical simulations in dimensions $d=1,2,$ and $3$, using binary disorder. Here, a fraction $c$ of the lattice sites are randomly labeled as 'defect' sites having a local reduction factor $w_n=w<1$ of the activation rate, while, for the rest of the sites, $w_n=1$. 
The simulation then goes as follows. Initially, each site is set to be active with a probability $1/2$. An active site ($n$) is randomly picked and it is either made inactive with a probability $\frac{1}{1+w_n\lambda}$ or, with the complementary probability, $\frac{w_n\lambda}{1+w_n\lambda}$, one of its $2d$ neighbors is randomly selected and is activated provided it was inactive. Such an update is coupled with a time increment $\Delta t=1/N(t)$, where $N(t)$ is the actual number of active sites.   
A special case of binary disorder is $w=0$, in which the defect sites do not affect the dynamics on the rest of the sites, and this corresponds effectively to a diluted lattice. As discussed in section \ref{sec:model}, we restrict the process to the giant component in this case. 
We considered cubic lattices of typical linear sizes, in order, $L=10^6,5000$, and $500$ in dimensions $d=1,2$, and $3$. Periodic boundary condition was applied in all cases. 
We measured the fraction of persistent sites as a function of time, which was also averaged typically over $10-100$ different realizations of disorder. 
To estimate the critical point, we performed simulations started from a single active seed for different values of $\lambda$ and plotted the average number of active sites against the survival probability which must show a power-law dependence at the critical point \cite{vd}.  

\begin{figure}[ht]
\includegraphics[width=8.5cm]{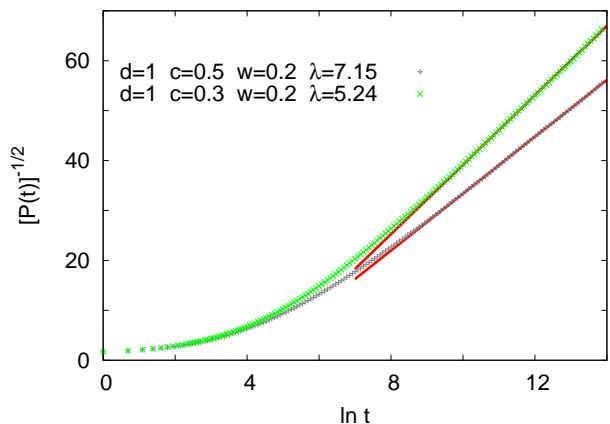}
\caption{\label{1d} (Color online) Dependence of the average persistence probability on time, obtained by numerical simulations in the one-dimensional, critical DCP for two different sets of parameters. According to Eq. (\ref{1d_bulk}), $[\overline{P}(t)]^{-1/2}$ must asymptotically increase linearly with $\ln t$. The straight lines are linear fits to the data.      
}
\end{figure}

\begin{figure}[ht]
\includegraphics[width=8.5cm]{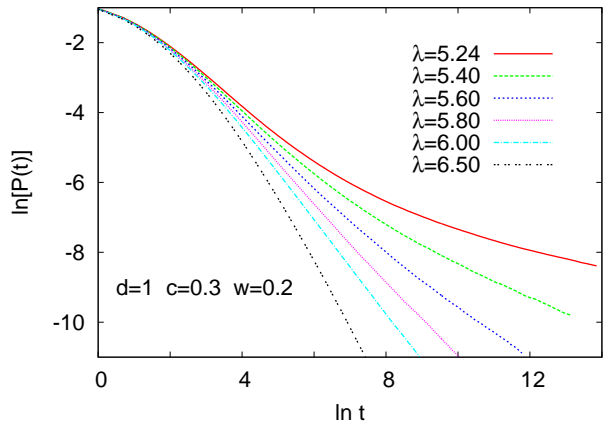}
\caption{\label{1dgp} (Color online) Dependence of the average persistence probability on time, obtained by numerical simulations in different points of the active phase of the one-dimensional DCP.   
}
\end{figure}

For the one-dimensional DCP, we considered two sets of parameters. For  $w=0.2$ and $c=0.3$ we made use of the estimate of the critical point $\lambda_c=5.24(1)$ from Ref. \cite{vd}, while, for $w=0.2$, $c=0.5$ we obtained $\lambda_c=7.15(5)$. 
As it is shown in Fig. \ref{1d}, the numerical results on the time-dependence of the average persistence are in accordance with the SDRG prediction in Eq. (\ref{1d_bulk}). In the active phase, the average persistence displays a power-law asymptotic decay with exponents varying with the control parameter, as can be seen in Fig. \ref{1dgp}, in agreement with the phenomenological result in Eq. (\ref{epl}).  

For the two-dimensional DCP, we considered again two sets of parameters. For $w=0.1$, $c=0.7$, we determined the critical point as $\lambda_c=5.085(5)$, while, for $w=0$, $c=0.2$, which corresponds to a diluted lattice, we used the estimate $\lambda_c=2.1075(1)$ from Ref. \cite{vfm}.  
At the critical point, we measured the average persistence, and fitted the function in Eq. (\ref{P_gen}) to the data (excluding the transient $\ln t<8$). To estimate the error of $\overline{\Theta}$, we plotted $[P(t)]^{-1/\overline{\Theta}}$ against $\ln t$ and determined the range of $\overline{\Theta}$ for which the asymptotic dependence is judged to be linear. 
This way, we obtained the estimates $\overline{\Theta}=0.73(4)$ for $w=0.1$, $c=0.7$, and $\overline{\Theta}=0.78(5)$ for the diluted lattice. 
These are somewhat higher than those obtained by the SDRG method, see Table \ref{table}, in which the estimates obtained by the two methods are summarized.
\begin{table}[h]
\begin{center}
\begin{tabular}{|r||r|r|r|}
\hline  $d$ & SDRG & MC ($w>0$) &  MC ($w=0$)  \\
\hline 
\hline  2   & 0.67(5) & 0.73(4)   &  0.78(5)           \\
\hline  3   & 0.33(6) & 0.34(4)   &  0.29(4)           \\
\hline
\end{tabular}  
\end{center}
\caption{\label{table} Numerical estimates of the generalized persistence exponent $\overline{\Theta}$ obtained by the SDRG method and Monte Carlo simulations  in dimensions $d=2$ and $d=3$.}
\end{table}
However, owing to the uncertainty of the estimation of $\lambda_c$ and corrections to the asymptotic form in Eq. (\ref{P_gen}), the true error of $\overline{\Theta}$ must be larger. This can be made visible in $d=1$, where $\overline{\Theta}$ is available analytically by the SDRG method.  
Fitting here the function in Eq. (\ref{P_gen}) in the same way to the MC data, the deviation of $\overline{\Theta}$ from the analytic value can be in the same order of magnitude as the observed difference between the MC and SDRG estimates in $d=2$. We conclude therefore that, in spite of the deviations, the MC estimates in $d=2$ are compatible with those of the numerical SDRG method.
\begin{figure}[ht]
\includegraphics[width=8.5cm]{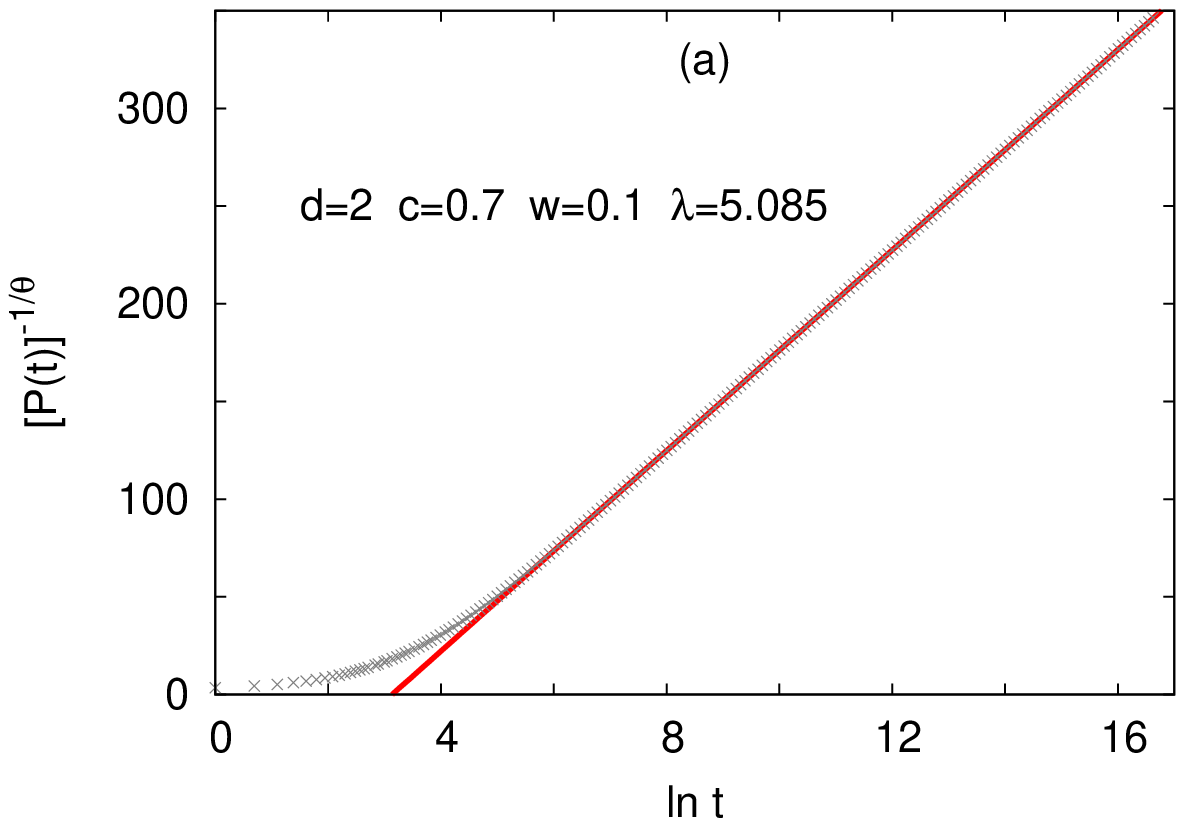}
\includegraphics[width=8.5cm]{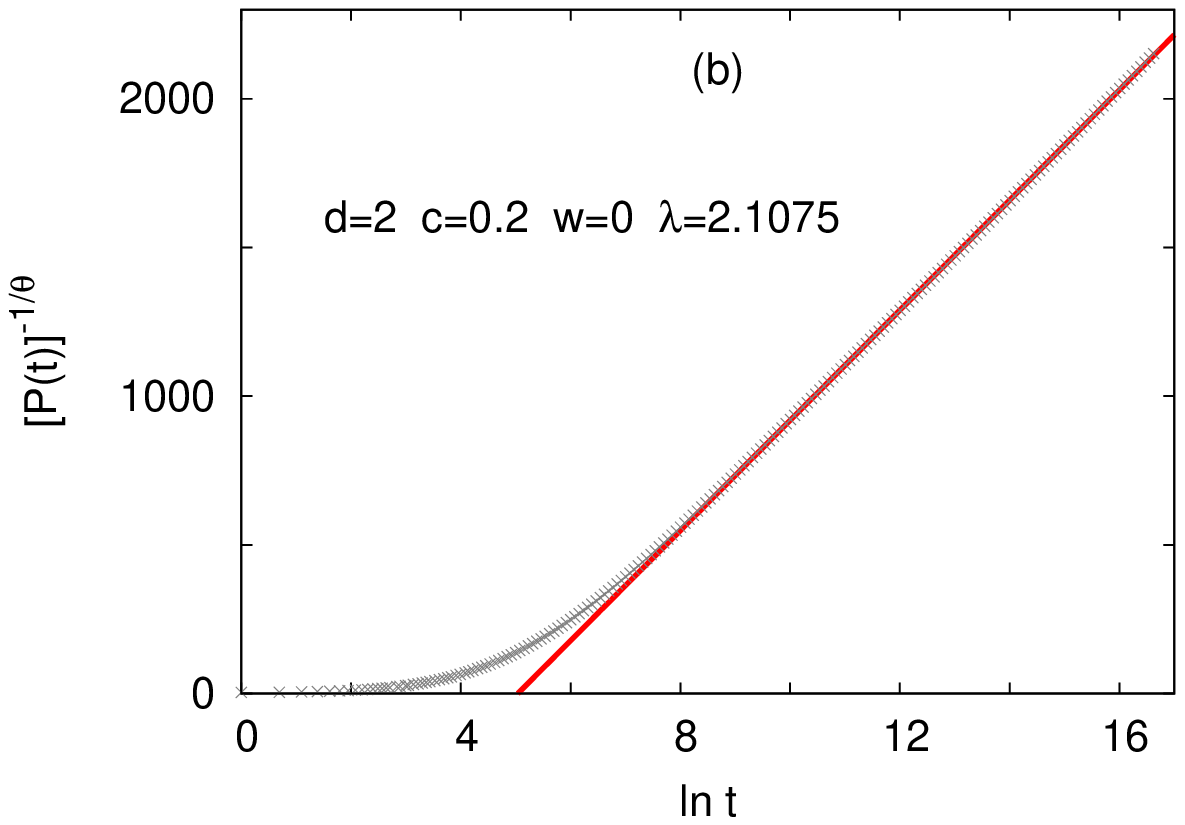}
\caption{\label{2d} (Color online) Dependence of the average persistence probability on time, obtained by numerical simulations in the two-dimensional, critical DCP for two different sets of parameters. The generalized persistence exponents used here are $\overline{\Theta}=0.73$ (a) and $\overline{\Theta}=0.78$ (b). The straight lines are linear fits to the data.      
}
\end{figure}

\begin{figure}[ht]
\includegraphics[width=8.5cm]{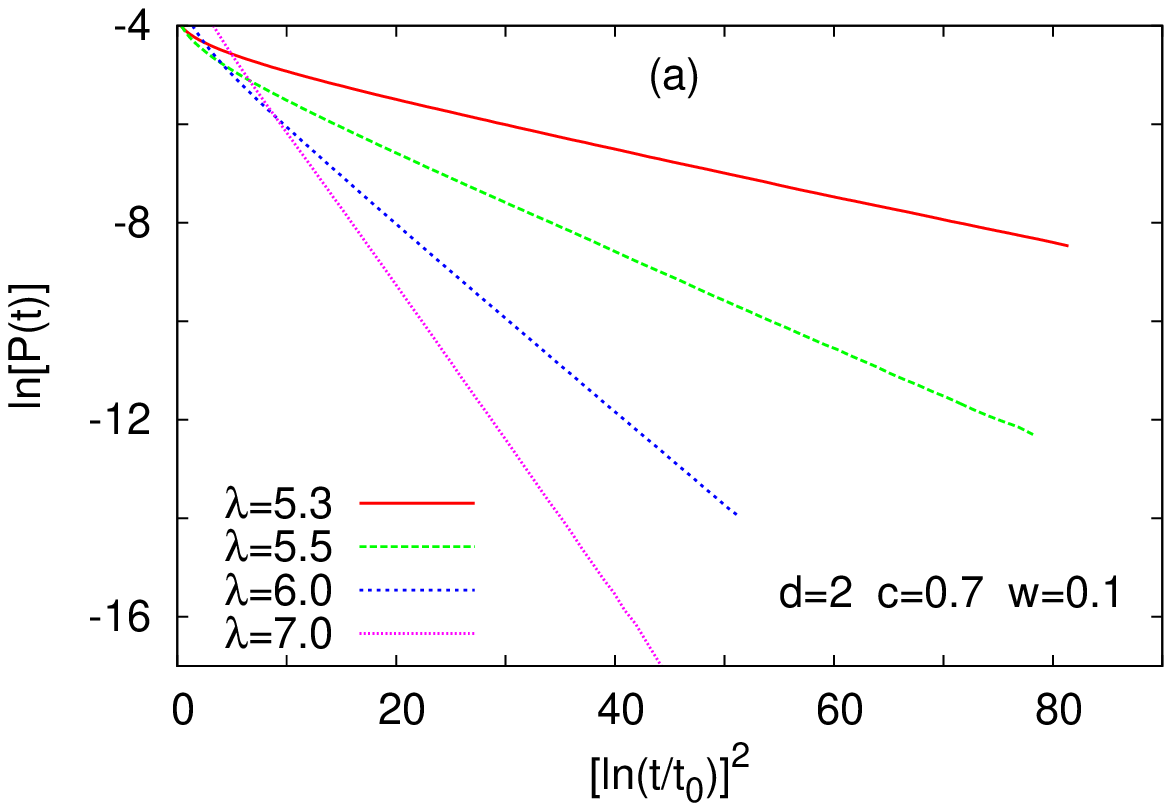}
\includegraphics[width=8.5cm]{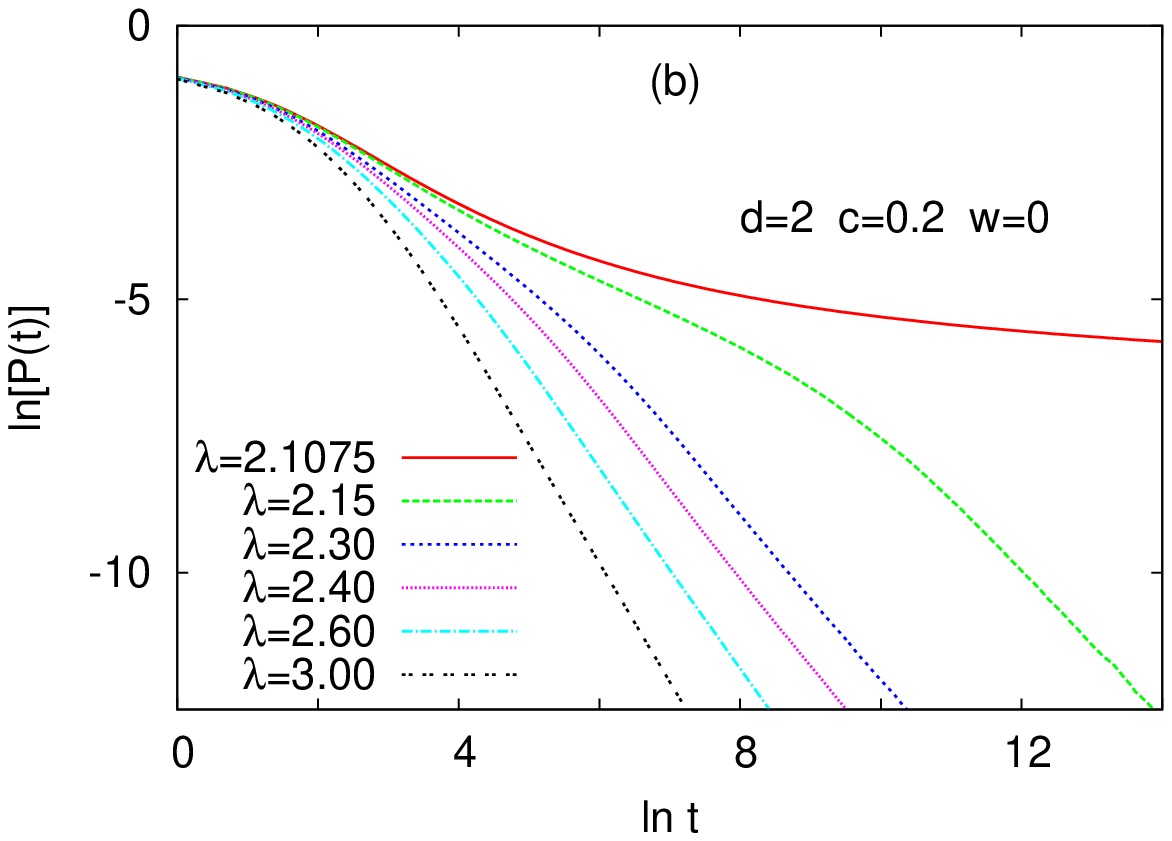}
\caption{\label{2dgp} (Color online) Dependence of the average persistence probability on time, obtained by numerical simulations in different points of the active phase of the two-dimensional DCP with parameters $w=0.1$, $c=0.7$ (a) and $w=0$, $c=0.2$ (b). In the top panel, the time scales are, for increasing $\lambda$, $t_0=1000,300,50$, and $10$.   
}
\end{figure}

The time-dependence of the average persistence in different points of the active phase is shown in Fig. \ref{2dgp} for the case $w=0.1$ and $c=0.7$. 
As can be seen, the numerical results support the enhanced-power law decay obtained by the phenomenological scaling considerations, see Eq. (\ref{epl}). 
As opposed to this, in the diluted lattice, the average persistence follows an algebraic decay given in Eq. (\ref{pl}) rather than an enhanced power law, as shown in Fig. \ref{2dgp}. Unlike in the one-dimensional model, which also displays an algebraic decay, the decay exponents seem to approach a non-zero limit as $\lambda$ tends to $\lambda_c$. This phenomenon will be explained in Sec. \ref{sec:discussion}.  

\begin{figure}[ht]
\includegraphics[width=8.5cm]{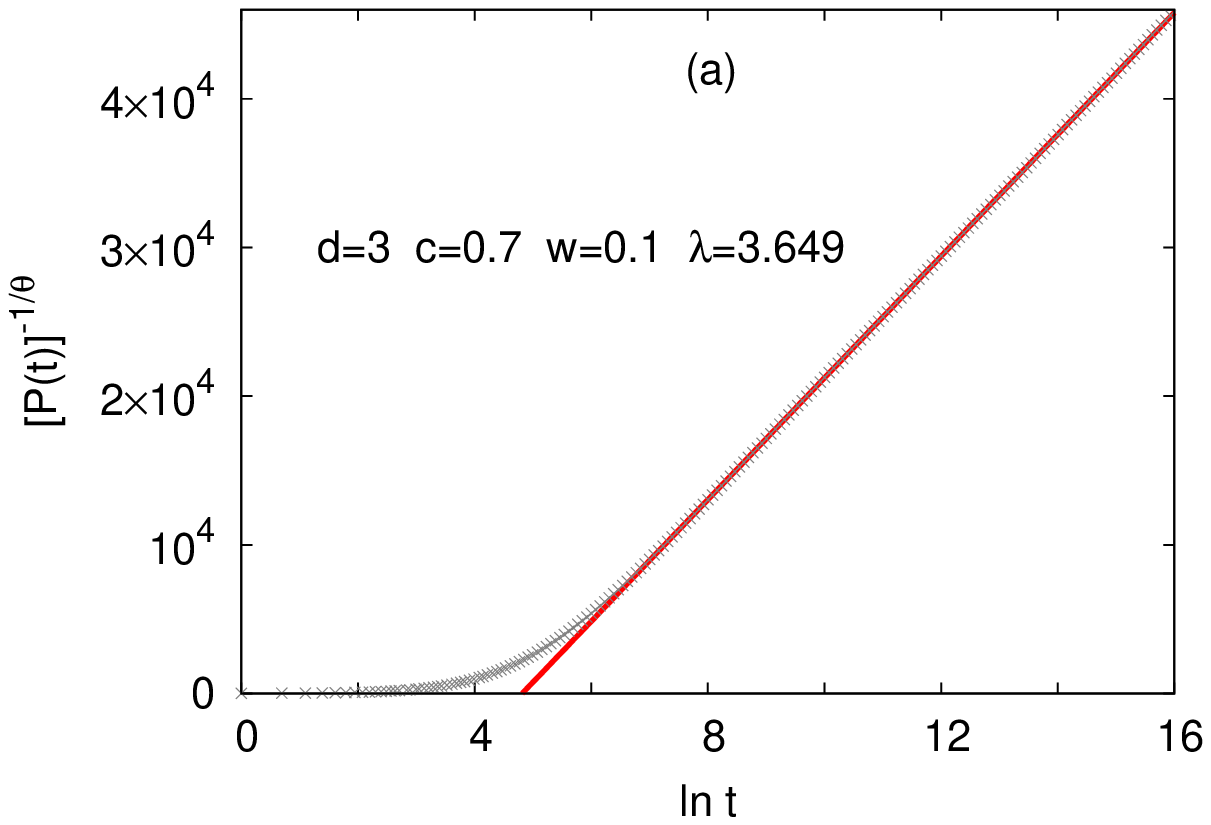}
\includegraphics[width=8.5cm]{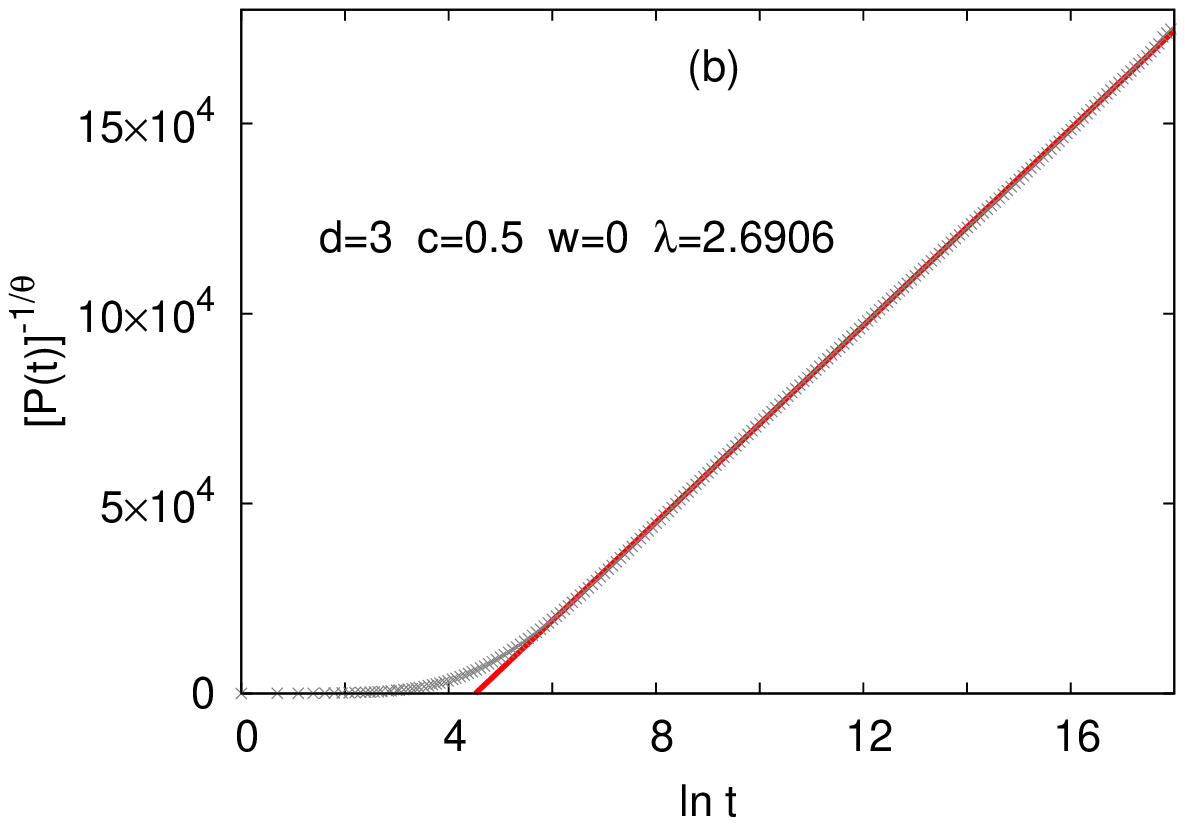}
\caption{\label{3d} (Color online) Dependence of the average persistence probability on time, obtained by numerical simulations in the three-dimensional, critical DCP for two different sets of parameters. The generalized persistence exponents used here are $\overline{\Theta}=0.34$ (a) and $\overline{\Theta}=0.29$ (b). The straight lines are linear fits to the data.      
}
\end{figure}

For the three-dimensional contact process we considered the parameter sets 
$w=0.1$, $c=0.7$, and $w=0$, $c=0.5$. In the former case, we obtained the estimate $\lambda_c=3.649(3)$, while in the latter case, which corresponds to a diluted lattice, we have taken the estimate $\lambda_c=2.6906(3)$ from Ref. \cite{vojta_3d}. At the critical point, the average persistence is found to follow the logarithmic law given in Eq. (\ref{P_gen}), and the generalized persistence exponents are estimated in the two cases to be $\overline{\Theta}=0.34(4)$ and $\overline{\Theta}=0.29(4)$, see Fig. \ref{3d}. These are again compatible with the estimates obtained by the numerical SDRG method, see Table \ref{table}. 

\begin{figure}[ht]
\includegraphics[width=8.5cm]{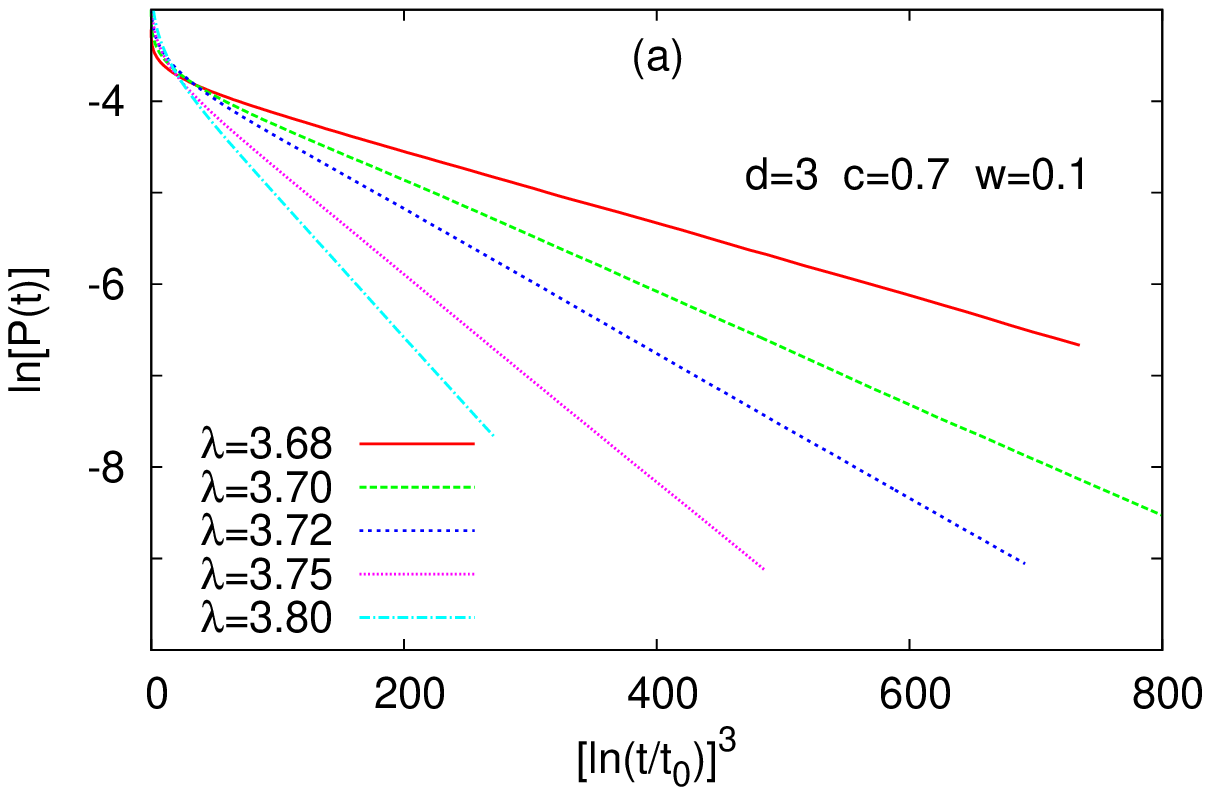}
\includegraphics[width=8.5cm]{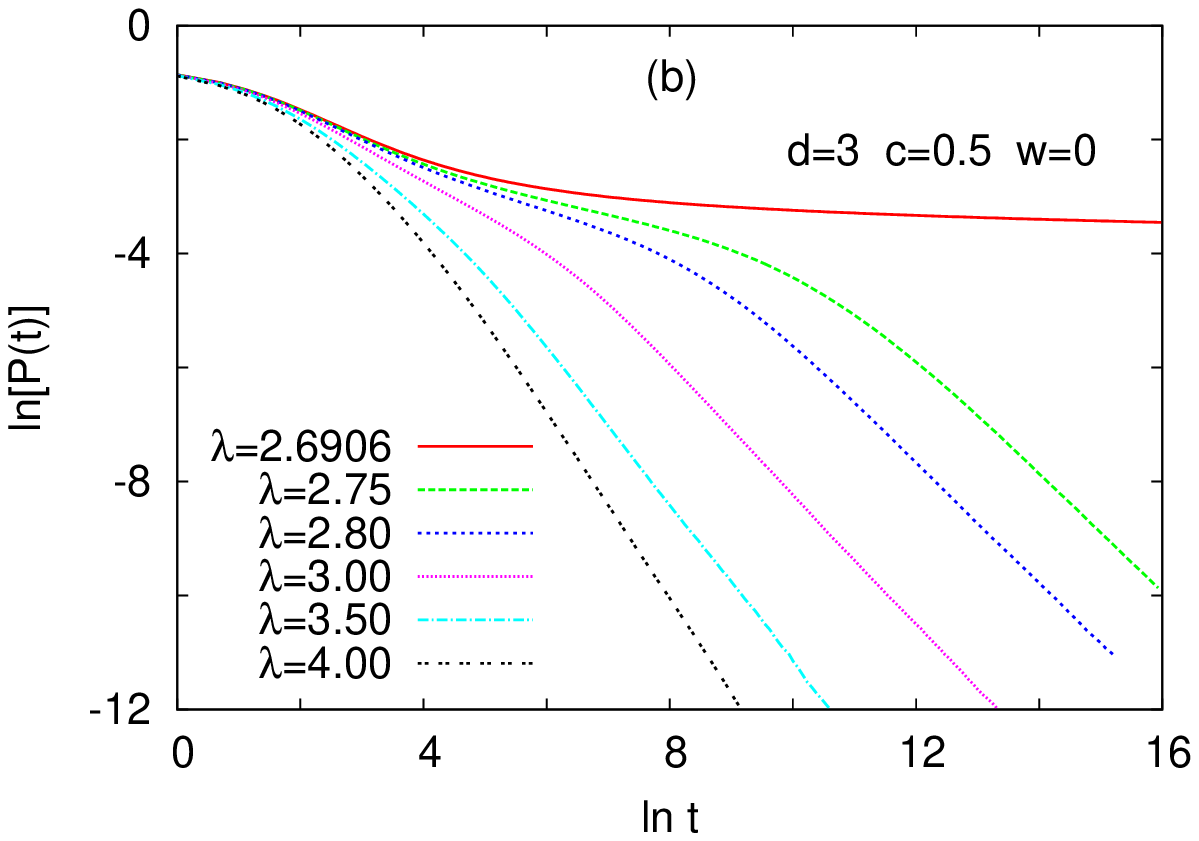}
\caption{\label{3dgp} (Color online) Dependence of the average persistence probability on time, obtained by numerical simulations in different points of the active phase of the three-dimensional DCP with parameters $w=0.1$, $c=0.7$ (a) and $w=0$, $c=0.5$ (b). In the top panel, the time scales are, for increasing $\lambda$, $t_0=1000,300,150,100$, and $50$.   
}
\end{figure}

The numerical results obtained in the active phase are similar to those obtained for $d=2$. As shown in Fig. \ref{3dgp}, the average persistence in the model with $w>0$ follows an enhanced power law, in accordance with Eq. (\ref{epl}). 
In the case of the diluted lattice ($w=0$), however, the average persistence decreases algebraically and the decay exponent seems to tend a non-zero limit as $\lambda\to\lambda_c$. 

\section{Discussion}
\label{sec:discussion}

In this work, we studied the time-dependence of the local persistence in the DCP during the evolution from a non-stationary initial state. 
We developped a method for calculating the average persistence in this system by the SDRG technique. We have found that the average persistence decays at the critical point at late times as an inverse power of $\ln t$, and determined the universal, dimension-dependent exponent analytically in one dimension and numerically in two and three dimensions.  
In one dimension we went beyond the calculation of the average and determined the limit distribution of the sample-dependent local persistences. According to the results, the persistence at late times can be characterized by a distribution of effective persistence exponents. 
In fact, the scheme for calculating the sample-dependent persistence formulated in the one-dimensional model can be easily generalized to higher dimensions. 
Observing the renormalization rule of persistence in Eq. (\ref{p_rg}), it turns out to be similar to the renormalization of the deactivation rate of a cluster when another cluster is merged with it, see Eq. (\ref{mu_rule}), ignoring the constant term.
Therefore the persistence of site $0$ in a given random sample can be calculated generally in the following way. Initially, the deactivation rate $\mu_0$ is set to a very small but non-zero value, which enables the calculation of persistence until the cluster containing site $0$ is decimated, at least down to a scale $\Omega=\mu_0$. At some rate scale $\Omega$ (with the above restriction), the effective deactivation rate $\tilde \mu_0$ of the cluster containing site $0$ is thus related to the variable $p$ describing persistence as $\tilde\mu_0(\Omega)=\tilde p(\Omega)\mu_0$. Therefore, the persistence probability at time $t=1/\Omega$ is given by $\tilde\mu_0(\Omega)/\mu_0$.      

In the active phase of the model, the average persistence decays anomalously due to rare-region effects, which is predicted to obey an enhanced power law by simple phenomenological arguments.  This behavior is confirmed by Monte Carlo simulations in non-diluted random systems. Yet, on the giant component of a percolating lattice the simulations show a power law decay. 
We attempt to explain this discrepancy in a phenomenological manner as follows.
Concerning the giant component of a diluted lattice, the average persistence can be decomposed into two contributions. One of them comes from the 'dangling ends' (DE) of the giant component \cite{bh}. These are small parts connected to the remaining part ('backbone') by a single path. Persistent sites within a DE, once the whole DE got into an inactive state, are highly protected against activation, which can come from the backbone only through a single path. This situation is essentially the same as in one dimension, thus these sites give an algebraically decaying contribution to the average persistence. 
Besides dangling ends, 'standard' rare regions can also form in the backbone, as regions of high local dilution. The occurrence of these is exponentially improbable in their volume, just like in the case of non-diluted random systems, hence their contribution decays as an enhanced power law in the active phase. Therefore, this contribution of the backbone is suppressed by the more slowly decreasing, algebraic contribution of DEs. At the critical point, however, the dominance is reversed, as the backbone yields a contribution of $O[(\ln t)^{-\overline{\Theta}}]$, suppressing the contribution of DEs.
When $\lambda$ is decreased in the active phase, the decay exponent in the dominant contribution of DEs slowly decreases. Yet, approaching $\lambda_c$, it will not get arbitrary close to zero since the size of the rare regions, the DEs, does not diverge as they are determined by the structure of the giant component, which is fixed.     

In summary, we have seen that, unlike the standard order parameter (density) of the model in the inactive Griffiths phase, the average persistence in the active phase is sensitive to the form of disorder. In the inactive phase, the rare regions are domains in which the interactions are stronger than the average \cite{noest}, whereas in the active phase, the anomalous behavior is caused by weakly interacting, less accessible regions, as it was discussed quantitatively in section \ref{sec:phen}. In the case of a dilution type of disorder, the accessibility of certain rare regions can be catastrophically low, altering the 'standard' behavior of persistence observed in non-diluted random systems: The appearance of isolated components in diluted systems leads to a non-zero average persistence even in the active phase, and, filtering out this contribution, the dangling ends are still able to change the standard behavior in the active phase.

\begin{acknowledgments}
This work was supported by the National Research, Development and Innovation Office -- NKFIH under grant No. K128989. IAK was supported by the European Research Council Synergy grant No. 810115 - DYNASNET.
\end{acknowledgments}


\appendix
\section{Relationship to a return probability}
\label{app:return}

From duality of the contact process \cite{schutz,hv}, we prove an exact equivalence between the local persistence and a return probability in a slightly modified system. In the quantum Hamiltonian formalism \cite{hv}, the configurations of a system with $L$ sites are described by states
 $|\eta\rangle\equiv \bigotimes_{i=1}^L|\eta_i\rangle$, where $\eta_i=0,1$ correspond to inactive and active state at site $i$, respectively. The state of the system at time $t$, $|\psi(t)\rangle = \sum_{\eta}p_{\eta}(t)|\eta\rangle$ evolves according to the master equation 
\be 
\partial_t|\psi(t)\rangle=-H|\psi(t)\rangle, 
\ee
where the 'quantum' Hamiltonian is given by 
\beqn 
H=-\sum_{i}\mu_i(s_i^+-n_i) - \nonumber \\ 
-\sum_{\langle ij\rangle}\lambda_{ij}[n_i\otimes(s_{j}^--v_{j}) + (s_i^--v_i)\otimes n_{j}].
\label{hamiltonian}
\eeqn
Here, only the non-trivially acting parts of the operators have been written out, and the summation in the second term goes over neighboring sites.
 Using a representation $(1, 0)^T$ and $(0, 1)^T$ of the states $|0\rangle$ and $|1\rangle$, respectively, the local operators appearing in Eq. (\ref{hamiltonian}) are represented by the matrices 
\be 
v=
\begin{pmatrix}
1 & 0 \cr
0 & 0  \cr
\end{pmatrix}
\quad 
n={\bf 1}-v
\quad 
s^-=
\begin{pmatrix}
0 & 0 \cr
1 & 0  \cr
\end{pmatrix}
\quad 
s^+=[s^-]^T
\ee
As the persistence probability on site $0$ is independent of $\mu_0$, it can be chosen arbitrarily. Let us set it to zero, $\mu_0=0$, and denote the Hamiltonian of this modified process by $H_0$. 
Let us consider now the evolution of the modified process from the initial state $|N_0\rangle$, in which all but site $0$ are active. 
Obviously, the persistence probability $P_0(t)$ of site $0$ is related to the local density $\rho_0(t)$ on site $0$ at time $t$ in the modified process as
\be 
P_0(t)=1-\rho_0(t). 
\ee
The state at time $t$ in the modified process is 
$|\psi_0(t)\rangle=e^{-H_0t}|N_0\rangle$ and the local density at site $0$ can be written as
\be
\rho_0(t)=\langle s|n_0|\psi_0(t)\rangle=\langle s|n_0e^{-H_0t}|N_0\rangle,
\label{rhoH}
\ee 
where $|s\rangle=\sum_{\eta}|\eta\rangle$.
With the Hamiltonian $H_0$, a dual Hamiltonian $\tilde H_0$ can be associated via 
\be 
\tilde H_0^T=DH_0D^{-1},
\ee
where $D=\bigotimes_{i=1}^L(v_i+s_i^-+s_i^+)$. The dual process differs from the original one in that $\lambda_{ij}$ and $\lambda_{ji}$ are interchanged \cite{hv,juhasz}. For a symmetric process (where $\lambda_{ij}=\lambda_{ji}$), the Hamiltonian is therefore self-dual, $\tilde H_0=H_0$.
Inserting the identity $D^{-1}D$ in Eq. (\ref{rhoH}) and using the relations 
$d_in_id_i^{-1}=v_i-s_i^+$, $\langle s|D^{-1}=\langle \varnothing |$, and $D|N_0\rangle = |\varnothing\rangle+|1_0\rangle$, where $|\varnothing\rangle$ and $|1_0\rangle$ denote the fully inactive state and the state with only site $0$ active, respectively, we obtain
\beqn
\rho_0(t)=\langle\varnothing|(v_0-s_0^+)e^{-\tilde H_0^Tt}(|\varnothing\rangle+|1_0\rangle)= \nonumber \\
=(\langle\varnothing|-\langle 1_0|)e^{-\tilde H_0^Tt}(|\varnothing\rangle+|1_0\rangle)=  \nonumber \\
=1-\langle 1_0|e^{-\tilde H_0^Tt}|1_0\rangle. 
\eeqn
After transposing it and assuming that the process is self-dual, we have 
\be 
P_0(t)=1-\rho_0(t)=\langle 1_0|e^{-H_0t}|1_0\rangle.
\ee
The r.h.s. is nothing but the return probability $P_{\rm ret}^{(0)}(t)$ to the state with only site $0$ active in the modified process.


\end{document}